\documentclass[aps, pre, twocolumn]{revtex4-1}
\usepackage{amsmath,amssymb,bm}
\usepackage{graphicx}
\usepackage[position=t,singlelinecheck=off,justification=raggedright]{subfig}

\newcommand{\myrefcite}[1]{Ref.~[\onlinecite{#1}]}
\newcommand{\myeqref}[1]{Eq.~\eqref{#1}}
\newcommand{\mysecref}[1]{Sec.~\ref{sec:#1}}
\begin{document}
\title{Kernel method for corrections to scaling}
\author{Kenji Harada}
\affiliation{Graduate School of Informatics, Kyoto University, Kyoto 606-8501, Japan}
\date{\today}

\begin{abstract}
  Scaling analysis, in which one infers scaling exponents and a
  scaling function in a scaling law from given data, is a powerful
  tool for determining universal properties of critical phenomena in
  many fields of science.  However, there are corrections to scaling
  in many cases, and then the inference problem becomes ill-posed by
  an uncontrollable irrelevant scaling variable. We propose a new
  kernel method based on Gaussian process regression to fix this
  problem generally. We test the performance of the new kernel method
  for some example cases. In all cases, when the precision of the
  example data increases, inference results of the new kernel method
  correctly converge. Because there is no limitation in the new kernel
  method for the scaling function even with corrections to scaling,
  unlike in the conventional method, the new kernel method can be
  widely applied to real data in critical phenomena.
\end{abstract}
\maketitle

\section{Introduction}
Critical phenomena relate many fields of science. Because of a scale
invariant at a critical point, a scaling law exists and different
systems share a set of values of scaling exponents. This is a
universality of critical phenomena. The set of exponent values defines
a universality class of critical phenomena. The classification of
various critical phenomena has been extensively studied from the
viewpoint of universality.
Scaling laws exist not only in the thermodynamic limit but also in
finite-size systems. In particular, by using a finite-size scaling law
for finite-size systems, we have confirmed various universality
classes\cite{Cardy:1988FSS}. However, such studies have often suffered
from corrections to scaling.
For example, in the study of spin glass transition, we often
experience difficulties attributed to strong corrections to scaling,
because the size of the system calculated is very limited. In the
study of exotic quantum criticality (e.g. the deconfined quantum
criticality \cite{Senthil:2004ui}, which is beyond the
Landau-Ginzburg-Wilson paradigm), the existence of corrections to
scaling increases the difficulty in distinguishing exotic quantum
criticality from a weak first-order transition\cite{Sandvik:2007dt,
  Sandvik:2010hg, Kaul:2012gx, Chen:2013bc, Pujari:2013df, Block:2013bx, Harada:2013bd, Suzuki:2015uj}.
Corrections to scaling are caused by the existence of irrelevant
scaling variables which is generally predicted by
renormalization-group theory. Therefore, treating corrections to
scaling is an important issue in research on critical phenomena.

The purpose of this work is to present a new approach to treat
corrections to scaling. With or without corrections to scaling, the
most conventional method in the scaling analysis of critical phenomena
is the least-squares method. It is based on chi-square statistics to
infer scaling exponents and a scaling function from data. If we
propose a certain model function for the scaling, the least-squares
method has no limitation. However, because we usually do not know the
specific scaling function, we assume a polynomial as the model
function in the least-squares method (see
\myrefcite{Slevin:1999gm}). Unfortunately, since it is difficult to
set the degree of the polynomial so that it approximates the scaling
function in the range of data points, we have to limit the range of
data to a narrow region near a critical point (see Fig. 4 in
\myrefcite{Harada:2011js}). The Bayesian inference method
\cite{Harada:2011js} was recently proposed to resolve this
inconvenient problem; it is based on a kernel method as the Gaussian
process regression. Recent studies of various critical phenomena have
proved the effectiveness of the kernel method\cite{Obuchi:2013fz,
  Motoyama:2013in, Harada:2013bd, Tomita:2014bh, Nakamura:2014vs,
  Takabe:2014gw, Takahashi:2015ky, Suzuki:2015uj, Okubo:2015fo}.  In
this work, we extend the Bayesian inference method to the problem of
corrections to scaling. The new kernel method is flexible even with
corrections to scaling, because only the smoothness of the scaling
function related to relevant scaling variables is assumed. Thus, it
can be a promising tool for studying critical phenomena whose data
cannot be treated by using the conventional approach.

The remaining parts of this paper are organized as follows. In
\mysecref{Scaling}, we will give a brief introduction to corrections
to scaling in the scaling analysis. In \mysecref{Bayes}, we will first
introduce the Bayesian inference and the Gaussian process regression
for the scaling analysis. We will then consider the ill-posedness of
an inference problem in the scaling analysis with corrections to
scaling, and we will propose a new composite kernel method. In
\mysecref{Tests}, we will explain the details of the practical
procedure for the new kernel method. After that, we will report the
performance of the new kernel method for example data sets in
detail. In \mysecref{Summary}, we will summarize this work.

\section{Scaling analysis and ill-posedness by corrections to scaling}
\label{sec:Scaling}
In this paper, for the sake of simplicity, we consider a finite-size
scaling law with a relevant scaling variable and an irrelevant
one. Although the irrelevant scaling variable vanishes in the
thermodynamic limit, it has an effect in a finite-size region, causing
a correction to scaling in the finite-size scaling law as
\begin{equation}
   A(t, u, L) = L^{c_2} F[ t L^{c_1},  u L^{-c_3} ],
\label{eq:fss}
\end{equation}
where $L$ is the size of a system, and a scaling variable $t$ ($u$) is
relevant (irrelevant). Thus, $c_1$ and $c_3$ are positive.  The
universality class of critical phenomena is defined by the exponent
values of the relevant scaling variables. The scaling function in the
thermodynamic limit is $F[\cdot, 0]$, and the correction to scaling is
$F[ t L^{c_1}, u L^{-c_3} ] - F[ tL^{c_1}, 0]$.

The object of the scaling analysis is to determine the critical
exponents from a data set of $A(t, u, L)$ by using the finite-size
scaling law \myeqref{eq:fss}. Using the new rescaled variables
\begin{equation}
  \label{eq:new_variable}
  X_1 \equiv tL^{c_1},\  X_2 \equiv u L^{-c_3},\ 
  Y \equiv A / L^{c_2},  
\end{equation}
we can rewrite \myeqref{eq:fss} as
\begin{equation}
  \label{eq:rescaling}
  Y = F(X_1, X_2) \pm E,
\end{equation}
where $E\equiv \delta A/L^{c_2}$ and $\delta A$ is the data precision
of $A$.  Thus, the scaling analysis is a statistical inference of
critical exponents so that rescaled data points collapse on a surface $F$ with precision
$E$.

In general, we can know the value of a relevant scaling variable
$t$. For example, $t \equiv T-T_c$, where $T$ is a temperature and
$T_c$ is a critical temperature. However, we do not know that of an
irrelevant scaling variable $u$. If $u$ is constant and non-zero, the
right-hand side of \myeqref{eq:fss} can be rewritten as
$L^{c_2} F[ t L^{c_1}, u L^{-c_3} ] = L^{c_2} G[ t L^{c_1}, L] = H[ t,
L ].$
Finally, the trivial scaling function can be defined as
$H$. Therefore, the inference problem in the scaling analysis with a
correction to scaling is ill-posed.

\section{Bayesian inference for scaling analysis}
\label{sec:Bayes}
\subsection{Bayesian inference}
If there is no irrelevant scaling variable, the definition of scaling
function $F[X_1,0]$ has no ambiguity. We can safely apply a
statistical inference method to the scaling analysis. The most popular
method is based on the leaset-squares method. From
\myeqref{eq:rescaling}, we can assume that the difference between $Y$
and $F[X_1,0]$ obeys the Gaussian distribution with mean $0$ and
variance $E^2$. Thus,
\begin{equation}
  \label{eq:least_square}
  P(Y) = \frac{1}{\sqrt{2\pi E^2}}\exp\left(-\frac{(Y - F[X_1])^2}{2E^2}\right),
\end{equation}
where $F[X] \equiv F[X, 0]$, $Y=A(T,L)/L^{c_2}$, $X_1=(T-T_c)L^{c_1}$,
$E=\delta A(T,L)/L^{c_2}$. The joint probability distribution of all
points $\vec{Y} \equiv (Y(1), Y(2), \cdots)^t$ is written as
\begin{equation}
  \label{eq:jpd}
  \log P(\vec{Y}) = -\sum_i \frac{(Y(i) - F[X_1(i)])^2}{2E(i)^2}
 - \sum_i \frac{\log\left(2\pi E(i)^2\right)}{2}.
\end{equation}
In the least-squares method, we assume the explicit form of the
scaling function $F[x]$ as a parametric function with a parameter set
$\vec{a}\equiv(a_0, a_1, \cdots)^t$. The best value of $\vec{a}$ is
inferred by maximizing the first term in the right-hand of
\myeqref{eq:jpd}, because the second term does not depend on
$\vec{a}$.

From the view point of Bayesian inference, this procedure can be
derived as follows. The right-hand side of \eqref{eq:jpd} depends on
the parameter set $\vec{a}$. Thus, it can be regarded as a conditional
probability of $\vec{Y}$ for $\vec{a}$ as $P(\vec{Y} \vert \vec{a})$.
According to Bayes' theorem, a conditional probability of $\vec{a}$
for $\vec{Y}$ can be written as
\begin{equation}
  P(\vec{a} \vert \vec{Y}) = 
  \frac{P(\vec{Y}\vert \vec{a})P(\vec{a})}{P(\vec{Y})}.
  \label{eq:bayes-th}
\end{equation}
When we do not know a prior distribution $P(\vec{a})$, we
usually suppose it as uniform. Then,
\begin{equation}
  P( \vec{a} \vert \vec{Y}) \propto
  P(\vec{Y}\vert \vec{a}).
  \label{eq:bayes-th-simple}
\end{equation}
Therefore, we can infer the most probable value of $\vec{a}$ by simply
maximizing the right-hand side of \eqref{eq:jpd} in Bayesian
inference.  It is called maximum \textit{a posteriori} probability (MAP)
estimation in the field of Bayesian inference.  The least-squares
method is a MAP estimation with \myeqref{eq:jpd}.

In many cases, a regression function $F[\cdot]$ linearly depends on
parameters in $\vec{a}$: for example, polynomial. Then, we can use a
simple $\chi^2$ test to check the quality of fit. However, $F[X_1]$
also includes physical parameters as $(T_c, c1, c2)$
nonlinearly. Thus, we cannot simply apply a $\chi^2$ test to check
the quality of fit with physical parameters.

We notice that the Bayesian inference does not restrict the form of a
conditional probability as \myeqref{eq:jpd}. Therefore, we
can design a suitable conditional probability for the scaling analysis
as follows.

\subsection{Gaussian process regression}
As mentioned above, we often use the least-squares method for a
scaling analysis. However, because one has to assume the form of the
unknown scaling function as like a polynomial, this sets limits on the
data in a narrow region near a critical point. To resolve this
difficulty, we introduced the view point of Bayesian inference in a
scaling analysis in \myrefcite{Harada:2011js}. In particular, we
explored the use of Gaussian process regression in a scaling
analysis. Gaussian process regression relies on a kernel function that
defines the covariance of the data, so it is called a \textit{kernel
  method}. Regarding data points as a Gaussian process with a
covariance matrix $\Sigma$, the conditional probability in
\eqref{eq:jpd} is changed to
\begin{equation}
  \label{eq:gpr}
P(\vec{Y} \vert \vec{a}) = \frac{1}{\sqrt{|2\pi\Sigma|}}
\exp\left(-\frac{\vec{Y}^t\Sigma^{-1}\vec{Y}}{2}\right),
\end{equation}
where $\vec{a}$ is a parameter set and $(\Sigma)_{ij} \equiv k(i,
j)$.
The kernel function $k(\cdot,\cdot)$ ensures that $\Sigma$ is positive
definite. If we use a Gaussian kernel $k_G$, the smoothness of the
regression function, which is a scaling function in our case can be
represented by a few parameters. The Gaussian kernel function in
\myrefcite{Harada:2011js} is written as
\begin{equation}
  \label{eq:gk}
  k_G(i,j) \equiv  (E(i)^2 + \theta_2^2)\delta_{ij}
  + \theta_0^2 \exp\left[ - \frac{|X_1(i)- X_1(j)|^2}{2\theta_1^2}\right].
\end{equation}
Here, we introduce new hyper parameters as
$(\theta_0, \theta_1, \theta_2)$.  If $X_1(i)$ is far from $X_1(j)$,
the kernel function exponentially decays.  The characteristic length
scale is controlled by a parameter $\theta_1$.  Since the covariance
$(\Sigma)_{ij}$ represents a correlation between $Y(i)$ and $Y(j)$,
the parameter $\theta_1$ represents the local smoothness of a scaling
function. The Gaussian kernel function only restricts the local
smoothness of the scaling function. It does not restrict the global
shape of the scaling function as in the least-squares
method. Therefore, no limitation on the range of data is needed in the
kernel method. We notice that the number of hyper parameters is only 3
as $(\theta_0, \theta_1, \theta_2)$.  For the least-squares method, we
probably need many parameters in a regression function to fit data in
a wide range.

In practice, we can infer the best values of parameters
$\vec{a}=(T_c, c_1, c_2, \theta_0, \theta_1, \theta_2)^t$ by maximizing
the likelihood [the right-hand side of \myeqref{eq:gpr} with
\myeqref{eq:gk}]. We used the conventional nonlinear optimization
method to find a maximum point. In such algorithms, we need the
derivative of \myeqref{eq:gpr} for a parameter $a$. Then, we can use
the following formula:
\begin{eqnarray}
\frac{\partial \log P(\vec{Y}\vert\vec{a})}{\partial a} &=& -\frac{1}{2}\mathbf{Tr}\left(\Sigma^{-1}
\frac{\partial \Sigma}{\partial a}\right)\nonumber\\
&-& \vec{Y}^t \Sigma^{-1}\frac{\partial
  \vec{Y}}{\partial a}
+ \frac{1}{2}\vec{Y}^t \Sigma^{-1}
\frac{\partial\Sigma}{\partial a}  \Sigma^{-1}
\vec{Y}.
\label{eq:derivative}
\end{eqnarray}

Combining \myeqref{eq:bayes-th-simple} with \myeqref{eq:gpr}, we can
define the distribution function of parameters under given data
points. If we use a Monte Carlo method with the weight as
\myeqref{eq:gpr}, we can evaluate an average and a confidential
interval of a parameter under given data points.

In the Gaussian process regression, we can also infer the $Y$ value of
a new additional point $(X, Y)$. In fact, we assume that all data
points obey a Gaussian process. In other words, the joint probability
distribution of given data points and the new additional point
$(X, Y)$ is also a multivariate Gaussian distribution as
\myeqref{eq:gpr}. Thus, a conditional probability of $Y$ for given
data points and a parameter can be written by a Gaussian distribution
with mean $\mu(X)$ and variance $\sigma^{2}(X)$:
\begin{eqnarray}
  \label{eq:meanGPK}
  \mu(X) &\equiv& \vec{k}^t\Sigma^{-1}\vec{Y},\\
  \sigma^{2}(X) &\equiv& k(X, X)-\vec{k}^t\Sigma^{-1}\vec{k},
\end{eqnarray}
where $\vec{k} \equiv (k(X(1), X), k(X(2), X), \cdots)^t$.

\subsection{Composite kernel}
The Gaussian kernel method can be formally generalized to the scaling
analysis with an irrelevant scaling variable in \myeqref{eq:fss}.
Because the scaling function in \myeqref{eq:fss} smoothly depends on
scaling variables, it can be represented by a Gaussian kernel of a
two-dimensional space of $X_1$ and $X_2$. However, because of the
scaling function's ambiguity as discussed above, the representation
does not work. The assumption of smoothness of a scaling function does
not resolve this ambiguity.

To resolve it, we consider a Taylor expansion of a scaling function
$F[X_1,X_2]$ by an irrelevant rescaled variable $X_2$ at a point
$(X_1, 0)$ as
\begin{equation}
  F[X_1, X_2] =
  \sum_{n=0}^{\infty} f_n(X_1)(X_2)^n,
\label{eq:Taylor}
\end{equation}
where $f_n(X_1) \equiv \left.\frac{1}{n!}\frac{d^nF[x,y]
  }{dy^n}\right\vert_{(X_1,0)}$ and $f_0(\cdot)$ is a scaling function in the
thermodynamic limit. When we introduce the cutoff of the degree of
the Taylor expansion, the functional form of the irrelevant rescaled variable
$X_2$ is always a polynomial without ambiguity. In addition, the
function $f_n$ depends only on the relevant scaling
variable. Therefore, there is no ill-posedness in the inference
problem for the scaling function. If we maintain the dependence on the
relevant variable as the general functional form as $f_n$, then the
Gaussian process regression can be defined by a composite kernel
written as
\begin{align}
  &k_C(i, j) \equiv \delta_{ij} (E(i)^2 + \theta_2^2)\nonumber\\
  &+ \sum_{n=0}^{M} \theta_{n,0}^2 \exp\left[ - \frac{|X_1(i)- X_1(j)|^2}{2\theta_{n,1}^2}\right] \left[X_2(i)X_2(j)\right]^n,
  \label{eq:CK}
\end{align}
where $M$ is the cutoff of the Taylor expansion. Here, we introduce
new hyper parameters as $(\theta_{n,0}, \theta_{n,1})$ for
$0 \le n \le M$ and $\theta_2$. To derive the composite kernel, we
assume that variables $X_1$ and $X_2$ are statistically
independent. Since the kernel function represents the covariance
between two points $i$ and $j$, the total kernel is a simple product
of kernels for each variable: a Gaussian kernel for $f_n(X_1)$ and a
polynomial kernel for $X_2^n$. The Gaussian function part in
\myeqref{eq:CK} as
$\theta_{n,0}^2 \exp\left[ - \frac{|X_1(i)-
    X_1(j)|^2}{2\theta_{n,1}^2}\right]$
represents the local smoothness of $f_n$ without the assumption of the
global shape of $f_n$. Therefore, it does not set any limits on the
data near a critical point. We notice that the case of $M=0$ is the
original kernel without a correction to scaling in
\myrefcite{Harada:2011js} (\myeqref{eq:gk} in this paper). Thus, the
present approach for the correction to scaling is a systematical
extension of the previous kernel method of the scaling analysis.

\section{Performance of the composite kernel method}
\label{sec:Tests}
We will now test the performance of the composite kernel method.  In
this section, for the sake of simplicity, we set the cutoff of the
Taylor expansion as $M=1$. We simply call the case of $M=1$ the
composite kernel in this section.

At first, we will explain computational techniques for the kernel
method in 
Secs.~\ref{sec:Normalization}--\ref{sec:Num}.
After that, in Secs.~\ref{sec:data} and \ref{sec:Ising},
we will report the performance of the composite kernel method in
detail.

\subsection{Normalization of rescaled variables}
\label{sec:Normalization}

In Gaussian process regression, we first use a nonlinear optimization
to find a good starting point for Monte Carlo sampling. Although there
is no absolute scale for rescaled variables, we found that the
normalization of rescaled variables increases the numerical stability
of the optimization process at the first stage of Gaussian process
regression. In this work, we set the unit of length scale by the
largest system size as
\begin{align}
  X_1 &= (T-T_c) (L/L_{\max})^{c_1} / R_X,\nonumber\\
  X_2 &= (L/L_{\min})^{-c_3},\nonumber\\
  Y &= (A / (L/L_{\max})^{c_2} - Y_0)/R_Y,\nonumber\\
  E &= \delta A/ (L/L_{\max})^{c_2}/R_Y,
      \label{eq:nv}
\end{align}
where $L_{\max}$ and $L_{\min}$ are the largest and smallest system
sizes in a data set, respectively. The values of $X_1$ and $Y$ and $E$
for the largest system are independent of relevant scaling exponents
($c_1$ and $c_2$). The scaling factor $R_X$ is defined so that the
width of $X_1$ for the largest system is 2.  The scaling factor $R_Y$
and the shift parameter $Y_0$ are defined so that $Y$ for the largest
system is in $[-1:1]$. Since data of the largest system most likely
affect the inference of a scaling function, the normalization of
rescaled variables may separate the inference of a scaling function
from those of scaling exponents.

\subsection{Hybrid Monte Carlo sampling}
\label{sec:HMC}
In the inference process of averages and confidential intervals of
parameters, we use a hybrid Monte Carlo method (see a review in
\myrefcite{neal2011mcmc}) for normalized rescaled variables to construct
samples of parameters by using the likelihood of Gaussian process
regression. It is very effective for a Monte Carlo sampling of
continuous variables.

We consider the weight $W(\vec{q})$ of a sampling parameter
$\vec{q}$. For example, in Gaussian process regression for a scaling
analysis by the composite kernel, $\vec{q}$ means inferred parameters
as
$\vec{a}= (T_c, c_1, c_2, c_3, \theta_2, \theta_{0,0}, \theta_{0,1},
\cdots, \theta_{M, 0}, \theta_{M, 1})^t$.
$W(\vec{q})$ is written as the likelihood of \myeqref{eq:gpr}.

In the hybrid Monte Carlo sampling, we introduce an artificial
momentum $\vec{p}$. Then, the artificial Hamiltonian is written as
\begin{equation}
  \label{eq:Ham}
  H(\vec{q}, \vec{p}) = U(\vec{q}) + \frac{\vec{p}^2}{2m},
\end{equation}
where $U(\vec{q})=-\log W(\vec{q})$ and $m$ is an artificial mass. At
first, we set an initial $\vec{p}(0)$ by a random variable which obeys
a Gaussian distribution with mean $0$ and variance $m$, and
$\vec{q}(0)=\vec{q}$. We calculate an artificial time-evolution of
$(\vec{q}(t), \vec{p}(t))$ until an artificial time $T$ by using a
leap-flog integration scheme:
\begin{align}
  \label{eq:leapflog}
  &\vec{p}(t+\epsilon/2) = \vec{p}(t) - (\epsilon/2) \nabla U(\vec{q}(t)),\nonumber\\
  &\vec{q}(t+\epsilon) = \vec{q}(t) + (\epsilon/m)  \vec{p}(t+\epsilon/2),\nonumber\\
  &\vec{p}(t+\epsilon) = \vec{p}(t+\epsilon/2) - (\epsilon/2) \nabla U(\vec{q}(t+\epsilon)),
\end{align}
where $\epsilon$ is a discrete time step.  We accept $\vec{q}(T)$ as a
next sample with the probability
\begin{equation}
  \label{eq:accept}
  \min\left[1, \exp\left(-H(\vec{q}(T), \vec{p}(T))+H(\vec{q}(0), \vec{p}(0))\right)\right].
\end{equation}
If not, a sample $\vec{q}$ is remained. Since the leap-flog
integration conserves the total energy, the acceptance probability is
almost 1. Therefore, the performance is much more effective than that of
the conventional approach using random walks to make trail samples. In
fact, we have to choose a suitable $m$ and $T$. In many cases, we need
to introduce individual masses. The practical details are explained in
\myrefcite{neal2011mcmc}.

\subsection{Practical procedure of the kernel method}
\label{sec:Num}
In this subsection, we summarize a practical calculation procedure in
our kernel method for a scaling analysis. It is based on the weight of
parameters $\vec{q}(=\vec{a})$ as
\begin{equation}
  \label{eq:gpr2}
W(\vec{q}) \equiv \frac{1}{\sqrt{|2\pi\Sigma|}}
\exp\left(-\frac{\vec{Y}^t\Sigma^{-1}\vec{Y}}{2}\right),
\end{equation}
where $(\Sigma)_{ij} \equiv k_C(i, j)$ and the composite kernel
$k_C(i,j)$ is defined by \myeqref{eq:CK}. The inferred parameter is
$\vec{q} = (T_c, c_1, c_2, c_3, \theta_2, \theta_{0,0}, \theta_{0,1},
\cdots, \theta_{M,0}, \theta_{M,1})^t$.
In addition, we use normalized rescaled variables in
\myeqref{eq:nv}. We estimate an average and a confidential interval of
parameters from this weight under a given data set by the hybrid Monte
Carlo sampling.

At first, to prepare a good starting point of Monte Carlo method, we
find a maximum point of a logarithmic weight $W(\vec{q})$ by using a
nonlinear optimization algorithm as the Fletcher-Reeves conjugate
gradient algorithm. The derivative of a weight by $\vec{q}$ is written
as
\begin{eqnarray}
\frac{\partial \log W(\vec{q})}{\partial q_i} &=& -\frac{1}{2}\mathbf{Tr}\left(\Sigma^{-1}
\frac{\partial \Sigma}{\partial q_i}\right)\nonumber\\
&-& \vec{Y}^t \Sigma^{-1}\frac{\partial
  \vec{Y}}{\partial q_i}
+ \frac{1}{2}\vec{Y}^t \Sigma^{-1}
\frac{\partial\Sigma}{\partial q_i}  \Sigma^{-1}
\vec{Y}.
\label{eq:derivative2}
\end{eqnarray}
The Cholesky decomposition of $\Sigma$ is useful to calculate the
right-hand side of \myeqref{eq:derivative2}. We notice that the
initial values of parameters are important, because a nonlinear
optimization scheme generally has no global convergence. In many
cases, we can guess the good initial values of physical parameters
($T_c, c_1, c_2, c_3$). We can also prepare the initial values of
hyper parameters ($\theta_2, \theta_{0,0}, \theta_{0,1}, \cdots, \theta_{M, 0}, \theta_{M, 1}$),
because we normalize the range of data points by introducing the
normalization of rescaled variables. In fact, we always start from 1
for all hyper parameters in the following cases. The hyper parameter
$\theta_2$ represents the total fidelity of a data set. Thus,
$\theta_2=1$ implies that all data points has no fidelity. In fact,
the values of parameters can widely move at the early stage of
non-linear optimization. It avoids trapping in local maximums. If the
precision of data is enough, the inference process will finally
converge with small $\theta_2$.

To infer averages and confidential intervals of parameters, we make
many samples of parameters by the weight in \myeqref{eq:gpr2}. We
recommend a hybrid Monte Carlo sampling to estimate an average and a
confidential interval of $\vec{q}$. The practical details of the
hybrid Monte Carlo method are reviewed in \myrefcite{neal2011mcmc}.

By using numerical libraries for the implementation of our kernel
method. we can easily make a code of our kernel method. The
performance tests in the following cases have been done by a reference
C++ code of our kernel method.\footnote{The reference code of our
  kernel method for the scaling analysis with or without a correction
  to scaling is available from
  \protect\url{http://kenjiharada.github.io/BSA/}.}

\subsection{Artificial data sets}
\label{sec:data}
\subsubsection{Dimensionless observable}
\begin{figure}[]
  \centering
  \subfloat[]{
    \includegraphics[width=0.44\textwidth]{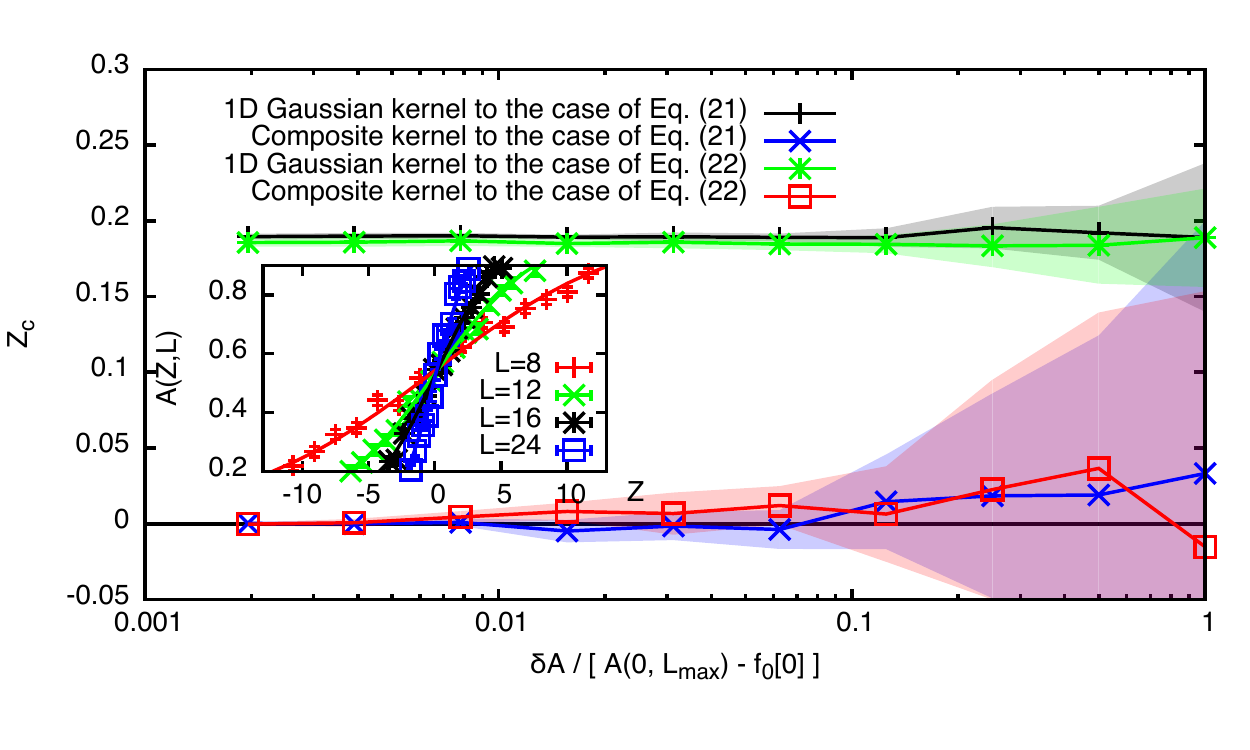}
  }\\
  \subfloat[]{
    \includegraphics[width=0.44\textwidth]{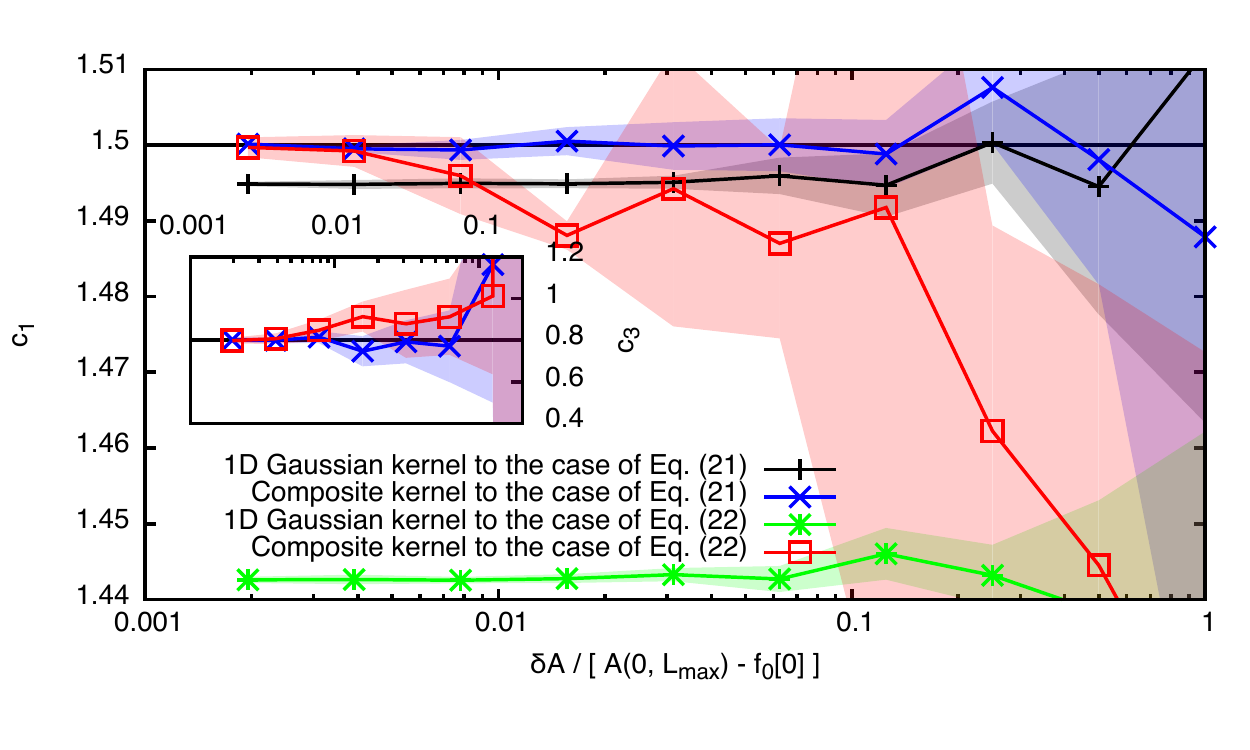}
  }\\
  \caption{(Color online) Average of inferred (a) $Z_c$, (b) $c_1$,
    and [inset of (b)] $c_3$ for ten data sets of $A$ with a precision
    $\delta A$. The shaded band denotes the variance of inferred
    results. Horizontal solid lines show correct values for
    each. Inset of (a): an example data set of $A$ obtained by using
    \myeqref{eq:A_C1}.  }
  \label{fig:C1}
\end{figure}
We first apply the composite and original kernel methods to two
artificial data sets with different types of correction forms. The
first data are defined as
\begin{equation}
  A(Z, L) \equiv f_0\left[(Z-Z_c)(L/L_0)^{c_1}\right] +a (L/L_0)^{-c_3},
\label{eq:A_C1}
\end{equation}
and the second data are defined as
\begin{equation}
A(Z, L) \equiv f_0\left[(Z-Z_c)(L/L_0)^{c_1}\right]\left(1 + \frac{a (L/L_0)^{-c_3}}{f_0[0]}\right),
\label{eq:A_C2}
\end{equation}
where $f_0[x]=\frac12 (\tanh x + 1)$, $t\equiv(Z-Z_c)$ is a relevant
scaling variable, $L$ is the system size. Both forms obey a
finite-size scaling law, and the $L^{-c_3}$ term is the correction to
scaling. In particular, both forms are equal to the first-order of the
Taylor expansion of \myeqref{eq:Taylor}: $f_0$ is shared, but $f_1$ is
constant or proportional to $f_0$, respectively. We choose the values
of the critical exponents close to those of the three-dimensional
Ising model; we set the values of parameters as $Z_c=0$, $c_1=1.5$,
$c_3=0.8$, $a=0.01$, and $L_0=48$. We generate an example data set of
$A$ by adding a Gaussian noise with mean 0 and variance $(\delta A)^2$
to \myeqref{eq:A_C1} or \myeqref{eq:A_C2}. We consider four different
system sizes: $L=8, 12, 16, 24.$ The number of data points for each
system size is 17. Because of the existence of the correction term,
the crossing point between different system sizes shifts from the
critical point [see the inset of Fig.~\ref{fig:C1}(a)]. The minimum
size of the correction term is about 4\% of the thermodynamic scaling
function at the critical point $Z=Z_c$.

Using kernel methods, we infer $Z_c$ and $c_1$ from a data set of $A$
with a precision $\delta A$. We start from correct values of physical
parameters $(Z_c, c_1, c_2, c_3)$ for the convergence of non-linear
optimization process in the first stage of our kernel method. Here we
fix $c_2$ as zero, because $A$ in \myeqref{eq:A_C1} or
\myeqref{eq:A_C2} is dimensionless. Because of the normalization of
rescaled variables, we can safely start from 1 for all hyper
parameters $(\theta_2, \theta_{i,0}, \theta_{i,1}).$ Starting from the
result of the first optimization step, we estimate averages and
confidential intervals of parameters by $1000$ hybrid Monte Carlo
samples.

Figures \ref{fig:C1}(a) and \ref{fig:C1}(b) show inference results for
$Z_c$ and $c_1$ as a function of precision $\delta A$,
respectively. We average out results of ten data sets.  We call the
original kernel in \myrefcite{Harada:2011js} as the one-dimensional
(1D) Gaussian kernel; it is equal to $M=0$ in \myeqref{eq:CK}.
Whereas inferences from the 1D Gaussian kernel quickly converge to
incorrect values, those from the composite kernel effectively converge
to the correct ones.  However, it is necessary for the data precision
to be within 10\% for the size of the correction term. The case of
\myeqref{eq:A_C2} may be harder than that of \myeqref{eq:A_C1},
because the deviation from the correct value of $c_1$ from the 1D
Gaussian kernel is larger. However, the composite kernel succeeds in
its inference from high-precision data in both cases without knowledge
of the correction form. We notice that the composite kernel has the
same performance as the 1D Gaussian kernel in the case of no
correction to scaling. In general, it is difficult to infer the value
of the irrelevant scaling exponent with precision. However, when the
data precision is improved, the result of the inference should
converge to a correct value. The inset of Fig.~\ref{fig:C1}(b) shows
inference results of irrelevant scaling exponents $c_3$ obtained by
using the composite kernel. Although the variances of the inferred
$c_3$ values are large in both cases, they effectively are improved by
the data precision. Finally, the inference results converge to a
correct value.
\subsubsection{General observable}
\begin{figure}[]
  \centering
  \includegraphics[width=0.46\textwidth]{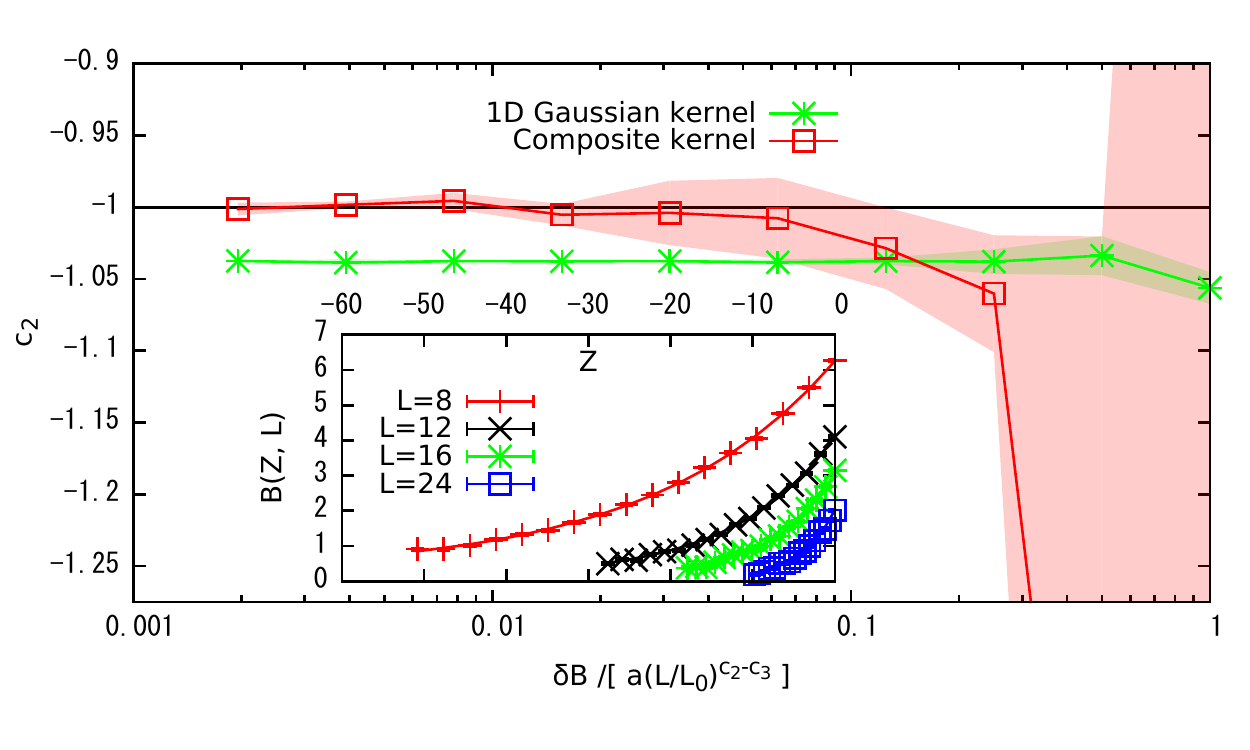}
  \caption{(Color online) Average of inferred $c_2$ for ten data sets
    of $B$ with a precision $\delta B$. The shaded band denotes the
    variance of inferred results. The horizontal line shows the
    correct value of $B$. Inset: an example data set of $B$.}
  \label{fig:C2}
\end{figure}

In the case of a general observable, we also need to infer
$c_2$ in the scaling law of \myeqref{eq:fss}. We test the case of a
general observable defined as 
\begin{equation}
  \label{eq:B}
\frac{B(Z, L)}{\left(\frac{L}{L_0}\right)^{c_2}} \equiv 
g_0\left[(Z_c-Z)\left(\frac{L}{L_0}\right)^{c_1}\right] + a \left(\frac{L}{L_0}\right)^{-c_3},
\end{equation}
where $g_0(x) = \exp(-c_2x/c_1)$ for $Z \le Z_c$. Here, we set $c_2$
as $-1$, and the values of other parameters are equal to those in the
dimensionless case of $A$. Figure \ref{fig:C2} shows the average of
inference results of $c_2$ as a function of precision $\delta B$. We
average out results of ten data sets of $B$. The composite kernel
succeeds in the inference of $c_2$ when the precision is within 10\%
for the size of the correction term. However, the 1D Gaussian kernel
method always converges to incorrect values, because it does not
consider the existence of a correction to scaling. We notice that the
performances for $Z_c$, $c_1$, and $c_3$ are similar to those in the
dimensionless case.

\subsection{Ising model on cubic lattices}
\label{sec:Ising}
Last, we apply kernel methods to the scaling analysis of the Ising
model on cubic lattices. The Hamiltonian is written as
$H \equiv -J\sum_{\langle ij \rangle} S_i S_j$, where the spin
variables are $S_i = \pm 1$, and $\langle ij \rangle$ denotes a pair of
nearest neighboring sites on a cubic lattice. The finite-temperature
phase transition defines the three-dimensional Ising universality
class, which widely covers a variety of critical systems. However, to
confirm the universality class precisely, we have to take into account
corrections to scaling in the scaling analysis
\cite{Hasenbusch:2010hv}. Fortunately, we can obtain high precision
data by using a sophisticated Monte Carlo algorithm for this
model. Thus, it is a good example for the scaling analysis with a
correction to scaling.

\subsubsection{Binder ratio of Ising model on cubic lattices}
We calculated Binder ratio,
$\langle (\sum_i S_i)^4 \rangle / [\langle (\sum_i S_i)^2 \rangle]^2$,
from $L=4$ to $32$ by using the cluster algorithm with an improved
estimator\cite{Swendsen:1987eq}. The simulation code is based on the
ALPS library\cite{Bauer:2011cp}. The number of temperature values for
each system size is 17.  To ensure high precision, we took about
$10^8$ samples for each point.

\begin{figure}[]
  \centering
  \subfloat[]{
    \includegraphics[width=0.44\textwidth]{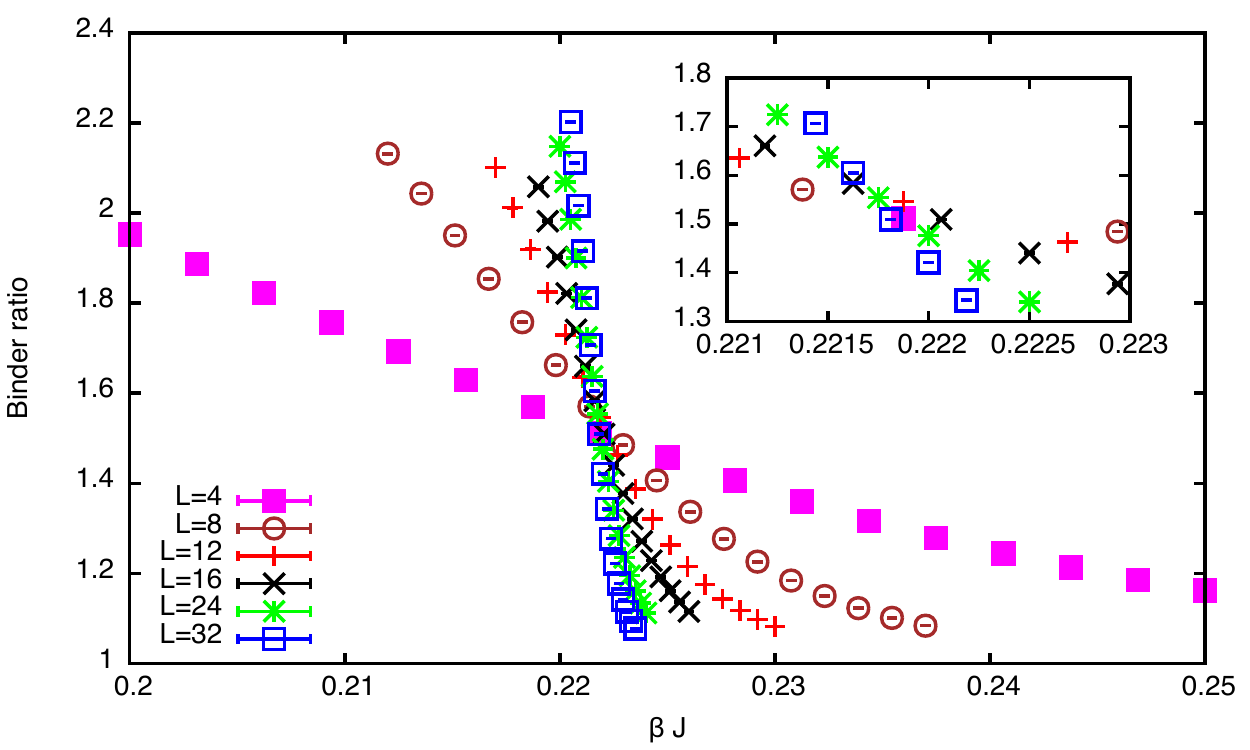}
  }\\
  \subfloat[]{
    \includegraphics[width=0.44\textwidth]{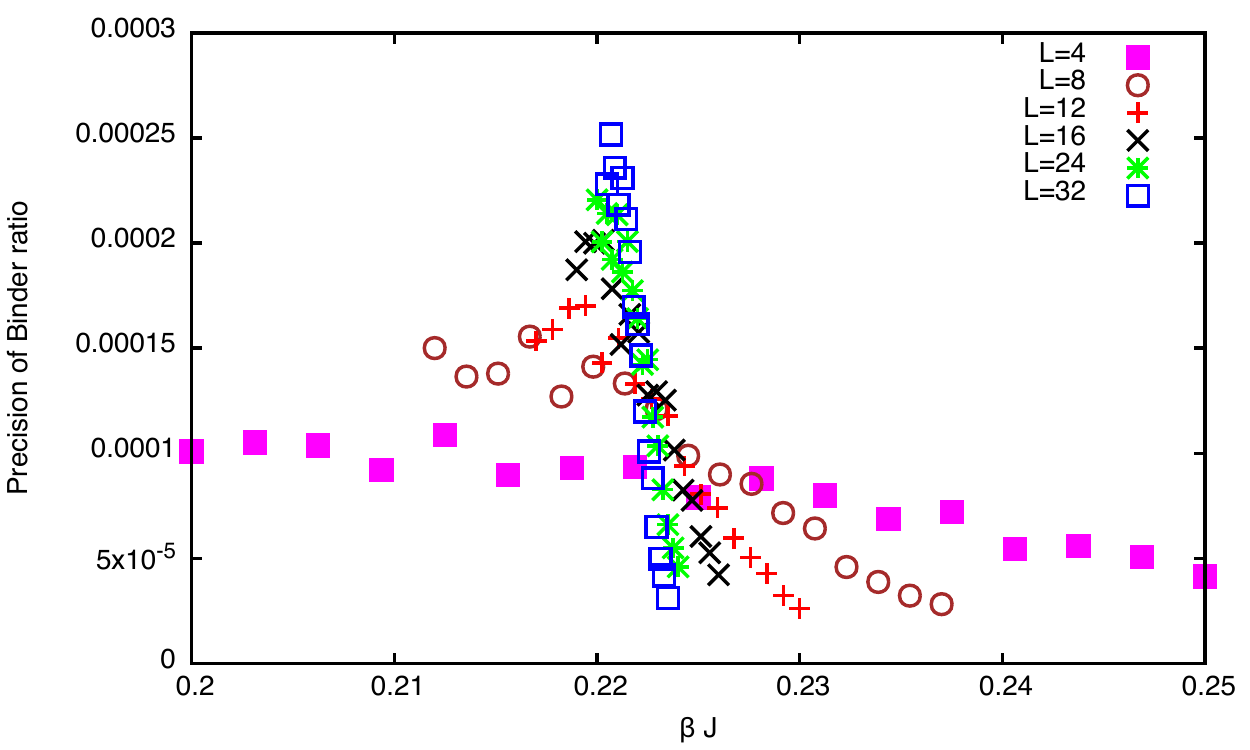}
  }
  \caption{(Color online) (a) Binder ratio of Ising model on cubic
    lattices. Inset: Binder ratio near a critical point. (b)
    Precision of Binder ratio in Monte Carlo calculations.}
  \label{fig:ap_1}
\end{figure}

Figure \ref{fig:ap_1}(a) shows Binder ratios from $L=4$ to $32$.  In the
thermodynamic limit, Binder ratios have a single crossing point at a
critical temperature. However, due to a correction to scaling, there
is a shift of crossing point in the finite system sizes.  As shown in
Fig.~\ref{fig:ap_1}(a), Binder ratios almost share a single crossing
point. The inset of Fig.~\ref{fig:ap_1}(a) shows Binder ratios near the
critical point [in \myrefcite{Hasenbusch:2010hv},
$\beta_c J = 0.22165463(8)$]. The crossing point between neighboring
system sizes shifts, indicating the existence of corrections to
scaling. Figure \ref{fig:ap_1}(b) shows the precision of data in Monte
Carlo calculations. Because of critical slowing down, the precision
becomes worse near a critical point. But, it may be enough to do a
scaling analysis with a correction to scaling.

\subsubsection{Inference results}
\begin{figure}[]
  \centering
  \subfloat[]{
    \includegraphics[width=0.46\textwidth]{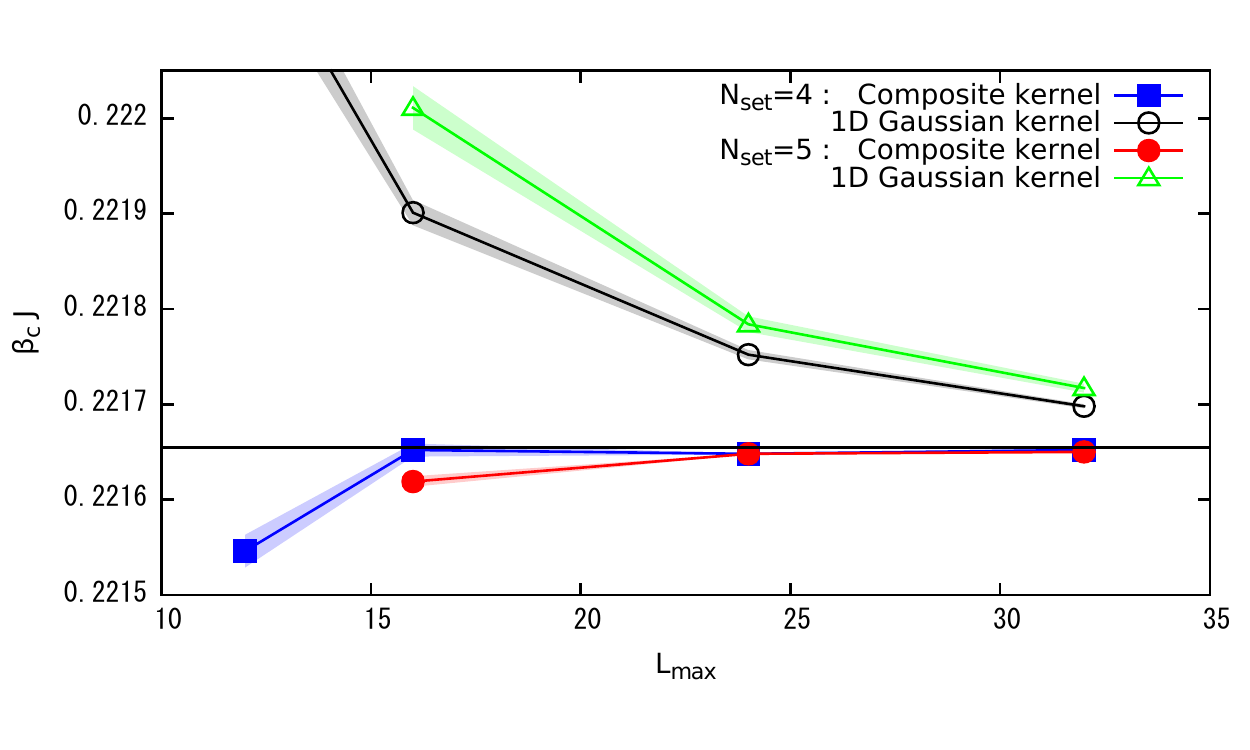}
  }\\
  \subfloat[]{
    \includegraphics[width=0.46\textwidth]{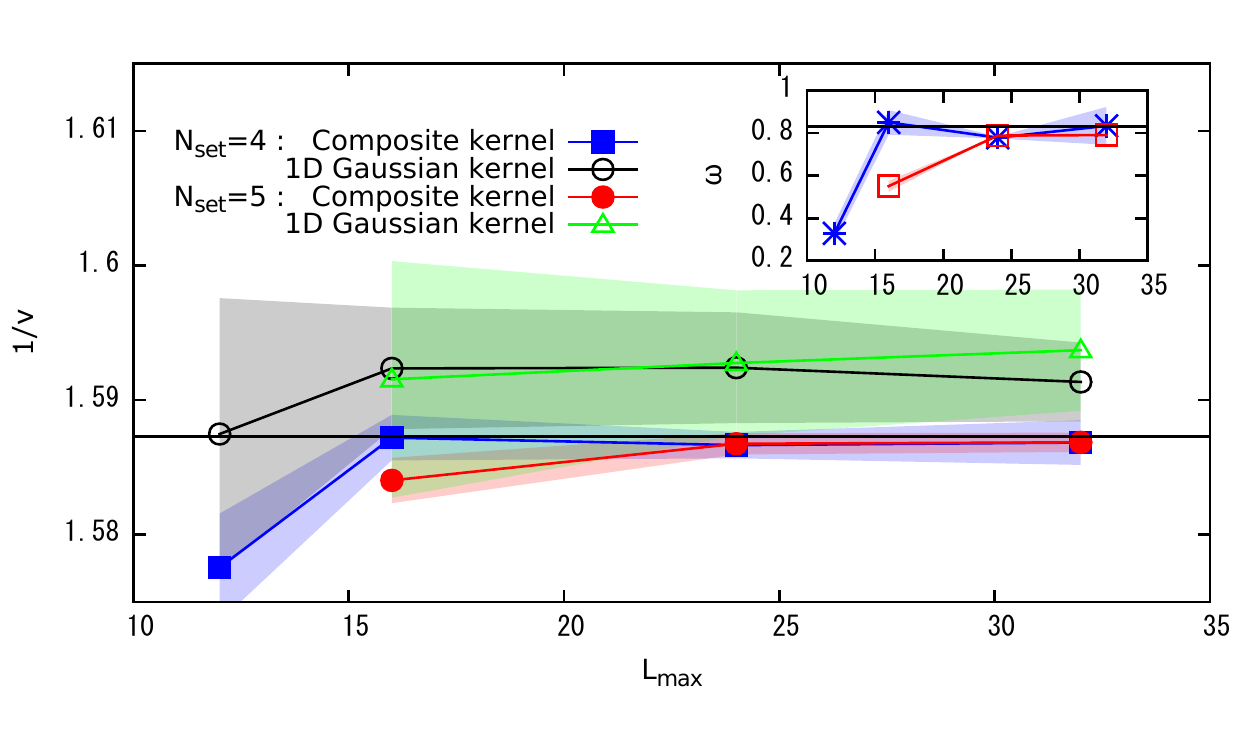}
  }
  \caption{(Color online) Inference results of (a) $\beta_c J$ and (b)
    $1/\nu$ from Binder ratios of the Ising model on cubic
    lattices. Inset of (b): Inference results of $\omega$.
    $N_{\rm set}$ is the number of systems in a data group to which
    kernel methods are applied. $L_{\max}$ is the largest system size
    in a data group. Horizontal solid lines show the values of
    $\beta_c J$, $1/\nu$, and $\omega$ in
    \myrefcite{Hasenbusch:2010hv}.  The shaded bands show the
    confidential intervals of inference results.}
  \label{fig:BR3D}
\end{figure}

\begin{figure}[]
  \centering
  \includegraphics[width=0.44\textwidth]{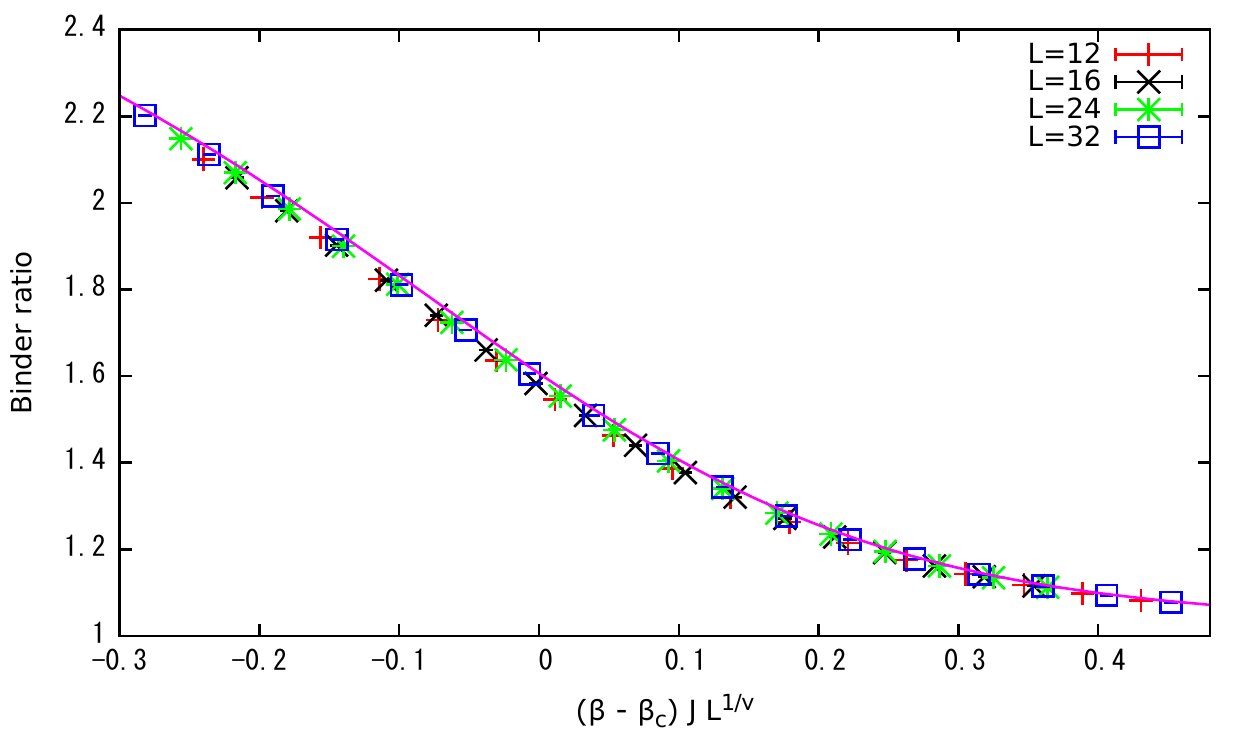}
  \caption{(Color online) Finite-size scaling plot of Binder ratios
    with a correction to scaling. The values of $\beta_c J$ and
    $1/\nu$ are 0.221652 and 1.587, respectively.  Solid pink line
    shows the thermodynamic scaling function inferred by the composite
    kernel.  }
  \label{fig:ap_3}
\end{figure}

We applied two kernel methods to the Binder ratios on selected system
sizes as a sequence of $N_{\rm set}$ systems up to $L_{\max}$. The
range of Binder ratios is $[1.1, 2.2]$. In addition, we start from
results of physical parameters in \myrefcite{Hasenbusch:2010hv} for
the non-linear optimization in the kernel method.  Figure
~\ref{fig:BR3D}(a) shows the inference results of an inverse critical
temperature $\beta_c J$ and a critical exponent $1/\nu$ as a function
of $L_{\max}$. The $\beta_c J$ results obtained by using the 1D
Gaussian kernel [$M=0$ in \myeqref{eq:CK}], in which a correction to
scaling is not assumed, slowly converge when $L_{\max}$ increases. In
contrast, results obtained by using the composite kernel [$M=1$ in
\myeqref{eq:CK}] quickly converge.  For critical exponents
$1/\nu$($=c_1$) and $\omega$($=c_3$), we observe a similar behavior in
Fig.~\ref{fig:BR3D}(b) and the inset. If the number of $N_{\rm set}$
increases, the ignorance of a correction to scaling affects an
inference result. In fact, the convergence of $N_{\rm set}=5$ is
slower than that of $N_{\rm set}=4$ in both kernel methods. The
$\beta_cJ$, $1/\nu$, and $\omega$ results obtained by using the
composite kernel for $N_{\rm set}=4$ and $L_{\max}=32$ are
$0.221652(2)$, $1.587(2)$, and $0.83(9)$, respectively. They are
consistent with reported values in \myrefcite{Hasenbusch:2010hv} of
$\beta_c J = 0.22165463(8)$, $1/\nu = 1.5873(6)$, and
$\omega=0.832(6)$, which were estimated from the fine-tuned
Blume-Capel model up to $L=360$. Figure \ref{fig:ap_3} shows the
finite-size scaling plot when we apply the composite kernel method to
the data set of $L=12, 16, 24$, and $32$ ($N_{\rm set}=4$ and
$L_{\max}=32$). Solid pink line shows the thermodynamic scaling
function inferred by the composite kernel. It is just
$\mu((X_1,X_2=0))$ in \myeqref{eq:meanGPK}. The size of the correction
to scaling can be roughly estimated from the inferred scaling function
and data. In this case, it is about 1\% of the scaling function near
the critical point, and the data precision is within 2\% for it.
Therefore, if the data precision is high enough, the composite kernel
can succeed in a scaling analysis of real data with a correction to
scaling.

\section{Summary}
\label{sec:Summary}
In this work, we proposed the composite kernel method for a scaling
analysis with corrections to scaling. This kernel has no ill-posedness
in the inference problem for the scaling analysis. The key to the new
kernel is the separation of relevant and irrelevant variables in the
inference problem. It is based on the Taylor expansion by an
irrelevant variable. We introduce the explicit form of corrections as
a polynomial of irrelevant variables. In contrast, we do not need the
explicit form of the scaling function related to the relevant
variables. Therefore, the composite kernel method can be widely
applied to real data in critical phenomena. We tested the performance
of the new kernel method for example data sets that have corrections
to scaling: three types of artificial data and a real data set of the
Ising model on cubic lattices. The new kernel succeeded in the scaling
analysis for all cases. In addition, we found that the data precision
is important for successful scaling analysis. If the data precision is
low, we cannot statistically distinguish a correction to scaling from
the data noise. A precision within 10\% for the correction term is
necessary for succeeding in the scaling analysis by using the
composite kernel method. Testing for a variety of critical phenomena
requires further studies.

The goodness of fit is a useful test in the least-squares method. In
the case of the kernel method, the regression function is non-linear
for inferred parameters. Thus, we cannot define a useful quantity
similar to the $\chi^2$. However, to be exact, the regression
function of the least-squares method is nonlinear for the inferred
physical parameters as $(T_c, c_1, c_2, c_3).$ Thus, a simple
$\chi^2$ test does not cover the check of their inferences. In
principle, the Bayesian framework gives us the distribution of
inferred parameters as a marginal likelihood. We can compare the
goodness of fit by the marginal likelihood. For example, we did not
discuss the cutoff of the Taylor expansion in \myeqref{eq:Taylor}. To
select a suitable cutoff, we can use the comparison of marginal
likelihoods. However, we need additional calculations of them with
addition of high precision for a high order correction term. The
simple procedure to check the goodness of fit remains in the future
study.

\appendix
\section*{Acknowledgments}
The author thanks Keith Slevin for enlightening discussions of the
least-squares method. The author also thanks Naoki Kawashima for
fruitful discussions. The present work is financially supported by
JSPS KAKENHI, Grant No. 26400392, Japan.

\bibliographystyle{apsrev4-1}
\bibliography{paper}

\end{document}